\documentclass[pra, aps, amsmath, amssymb, twocolumn, superscriptaddress]{revtex4-1}
\usepackage[utf8]{inputenc}
\usepackage[T1]{fontenc}
\usepackage{xcolor}
\usepackage{graphicx}
\usepackage{hyperref}
\usepackage{physics}
\usepackage{color}
\usepackage{ulem}
 \usepackage[caption=false]{subfig}

\begin{document}

\title{Entropy production  and asymptotic factorization via thermalization: a collisional model approach }
 %Joint state of a quantum system and its thermal bath: a conjecture beyond local thermalization}

\author{Stefano Cusumano}
\email[Corresponding author: ]{stefano.cusumano@sns.it}
\affiliation{NEST, Scuola Normale Superiore and Istituto Nanoscienze-CNR, I-56127 Pisa, Italy}

\author{Vasco Cavina}
\affiliation{NEST, Scuola Normale Superiore and Istituto Nanoscienze-CNR, I-56127 Pisa, Italy}
%\email{vasco.cavina@sns.it}

\author{Maximilian Keck}
\affiliation{NEST, Scuola Normale Superiore and Istituto Nanoscienze-CNR, I-56127 Pisa, Italy}
%\email{maximilian.keck@sns.it}

\author{Antonella De Pasquale}
\affiliation{Dipartimento di Fisica e Astronomia, Universit\'a di Firenze, I-50019, Sesto Fiorentino (FI), Italy}
\affiliation{INFN Sezione di Firenze, via G.Sansone 1, I-50019 Sesto Fiorentino (FI), Italy}
\affiliation{NEST, Scuola Normale Superiore and Istituto Nanoscienze-CNR, I-56127 Pisa, Italy}

%\email{antonella.depasquale@sns.it}

\author{Vittorio Giovannetti}
\affiliation{NEST, Scuola Normale Superiore and Istituto Nanoscienze-CNR, I-56127 Pisa, Italy}
%\email{vittorio.giovannetti@sns.it}

\begin{abstract} 

The Markovian evolution of an open quantum system is characterized by a positive entropy production, while the global entropy gets 
redistributed between the system and the environment degrees of freedom.
Starting from these premises, we analyze the entropy variation of an open quantum system in terms of two distinct relations:
the Clausius inequality, that provides an intrinsic bound for the entropy variation in terms of the heat absorbed by the system,
and an extrinsic inequality, which instead relates the former 
 to the corresponding  entropy increment of  the environment. 
By modeling the thermalization process with a Markovian collisional model, we compare and discuss the two bounds,
showing that the latter is asymptotically saturated in the limit of large interaction time.
In this regime not only the reduced density matrix of the system reaches 
an equilibrium configuration, but it also factorizes from the environment degrees of freedom.
This last result is proven analytically when the system-bath coupling is sufficiently strong 
and through numerical analysis in the
weak-coupling regime.
\end{abstract}

\maketitle

%%%%%%%%%%%%%%%%%%%%%%%%%%%%%%%%%%%%%%%%%

\section{Introduction} \label{SEC0} 
In recent times the research in thermodynamics has rose to new life, with special attention to the energy balance in nanoscale devices and a growing interest towards its connections to information theory~\cite{Goold2016}.
In this framework an issue naturally arises about the relations between thermodynamic and information theoretic quantities, for instance between the concept of  entropy introduced by Clausius, and the one introduced by Shannon~\cite{Shannon1948a, Shannon1948b}.
Such studies are complicated by the fact that for 
open quantum dynamical processes~\cite{BreuerPetruccione} 
a definition of heat and work for a quantum engine cannot be unequivocally given in the presence of correlations and non-negligible coupling
strengths \cite{Alipour2016, Weimer2008, Vinjanampathy2016,Lostaglio2015, Cwiklinski2015, Alicki1979, Talkner2007}, or more in general when the system is driven out 
of equilibrium \cite{Cavina2017}.

In this work we focus our attention on the  irreversible character of the thermalization process, by studying
how the entropy of a quantum system A  is modified  when it
is placed into thermal contact with an external reservoir.
Specifically, the entropy increase of A is analyzed in terms of two different inequalities: the Clausius formulation of the second law and one that accounts for creation of quantum and classical correlations between A and its environment.
%Specifically we are interested in comparing the bound on  the entropy increase of A deriving from the Clausius formulation of the second law 
%with the one that instead follows from general considerations on how the  degree of freedoms get redistributed among A and its environment. 
A detailed analysis of these relations is performed for the  thermalization scheme via collisional events 
 introduced by Scarani {\it et al.} in Ref.~\cite{Scarani2002}.  
Collisional Models (CMs)~\cite{first_coll_mod,BRUN,ZIM}
 have  been extensively used to study open quantum systems in a variety of situations, from cascade systems and networks~\cite{Palma2012,Giovannetti2012,Cusumano2017}, to heat 
transfer~\cite{Lorenzo2015b} and thermalization~\cite{Man2018,Scarani2002}. 
Apart from their intrinsic simplicity, they allow for exact tracking of the environmental degrees of freedom,  a  fundamental feature that 
lets one account the  entropy and energy balance in the system. 
Via exact analytic results extended by numerical analysis we get a new insight into the final state of the system and the bath after the thermalization process is completed: in particular we observe that for the scheme of Ref.~\cite{Scarani2002} the final state of  A not only reaches thermalization locally, but also gets completely factorized from the many-body environment
that triggers the thermalization via collisions (an effect we dub {\it asymptotic factorization} of A). 
 An explicit proof of this phenomenon
is given which works in the strong-collisions limit  regime. We conjecture that the same result should apply also in the 
weak-collisions limit and present numerical evidence of this fact for the special case where A is a qubit system.

The manuscript is organized as follows: 
in Sec.\ref{SEC1}, using a standard Hamiltonian characterization of the thermalization process, we present two different bounds on the entropy increase of A and clarify
their relations. In Sec.~\ref{SEC2} instead we review the CM of Ref.~\cite{Scarani2002} and study how the previous bounds affect the dynamics of the system
in this setting.  
Sec.~\ref{SUBSECNEW} is devoted to the study of the asymptotic factorization property. The paper ends in Sec.~\ref{SEC3}  where
we draw our conclusions. Technical derivations are reported in the Appendix. 

%%%%%%%%%%%%%%%%%%%%%%%%%%%%%%%%%%%%%%%%%%%%

  \section{Quantifying irreversibility} \label{SEC1}  
     Consider a quantum system A weakly coupled to a thermal environment B 
 at temperature $T$.
 Suppose that, due to its interaction with B, the system evolves 
 with
 an entropy increase $\Delta S_{\rm{A}}$ and absorbing an amount of heat $\Delta Q_{\rm{A}}$.
  Purely thermodynamic considerations suggest
     that 
     %if  the AB system is globally isolated,   then $\Delta Q_{\rm A}$ should correspond to  $-\Delta Q_{\rm B}$
%as a direct consequence of the energy conservation principle, and that 
   \begin{eqnarray}  \label{eq:Clausius}
\Delta S_{\rm A} \geq  \beta \Delta Q_{\rm A}\;, \quad  \mbox{(Clausius inequality)}
\end{eqnarray} 
as a consequence of the second law (hereafter we use $\beta= 1/(K_B T)$ to indicate the inverse temperature of the bath, $K_B$ being the Boltzmann constant). 
Equation~(\ref{eq:Clausius}) provides an "intrinsic" lower bound on the local entropy production as it involves only quantities 
that explicitly refer to properties of the system~A. By 
properly accounting the onset of classical and quantum correlations during the thermalization event,  an "extrinsic" bound relating $\Delta S_{\rm A}$ to the corresponding entropy increase of 
the bath $\Delta S_{\rm B}$ can also be obtained, leading to the inequality
  \begin{eqnarray} \Delta S_{\rm A} \geq - \Delta S_{\rm B}\;,
  \label{eq:Info} \end{eqnarray} 
  (more on this  in the following). 
Limiting how the entropy of A evolves,
Eqs.~(\ref{eq:Clausius}) and (\ref{eq:Info}) provide a characterization of  the irreversibility of the thermalization process. 
As we shall clarify later on they are not completely independent, 
even though no universal 
ordering between them can be established.
 A formal derivation of these results 
 can be obtained by modelling the AB coupling as a time-independent Hamiltonian  
     $\hat{H} = \hat{H}_{\rm A} + \hat{H}_{\rm B} + \hat{H}_{\rm int}$ and 
  assuming that no correlations are  shared between A and B before their interaction, i.e. 
writing  the initial state of the joint system as a factorized density matrix 
\begin{eqnarray} \hat{\rho}_{\rm AB}(0)=\hat{\rho}_{\rm A}(0) \otimes \hat{\eta}^{(\beta)}_{\rm B}\;,\label{INPUT} 
\end{eqnarray} 
 where $\hat{\rho}_{\rm A}(0)$ is the input 
state of A while 
$\hat{\eta}^{(\beta)}_{\rm B} := e^{-\beta \hat{H}_{\rm B}}/Z_{\rm{B}}(\beta)$  is the Gibbs density matrix describing the thermal equilibrium of the bath, $Z_{\rm{B}}(\beta) := \mbox{Tr}[ 
  e^{-\beta \hat{H}_{\rm B}}]$ being  the  partition function. 
With this specification 
the temporal evolution of A  can  now be fully specified  by 
the one-parameter family of Completely Positive,
 Trace (CPT) preserving channels $\{ \Phi_{0\rightarrow t}\}_{t\geq 0}$ describing, for arbitrary times $t$, the mapping
 \begin{eqnarray} \label{DEFMAP} 
 \hat{\rho}_{\rm A}(0) \rightarrow \hat{\rho}_{\rm A}(t) = \Phi_{0\rightarrow t}[ \hat{\rho}_{\rm A}(0)] := \mbox{Tr}_B[\hat{\rho}_{\rm AB}(t)  ]\;, \end{eqnarray} with 
 $\hat{\rho}_{\rm AB}(t) := e^{-i \hat{H}t/\hbar}(\hat{\rho}_{\rm A}(0) \otimes \hat{\eta}^{(\beta)}_{\rm B}) e^{i \hat{H}t /\hbar}$ 
being  the evolved state of the joint system.
We then say that B induces thermalization on A if,  irrespectively from the specific choice of $\hat{\rho}_{\rm A}(0)$,  the latter will be driven by $\Phi_{0\rightarrow t}$ into the equilibrium configuration state  
$\hat{\eta}^{(\beta)}_{\rm A}:= e^{-\beta \hat{H}_{\rm A}}/Z_{\rm{A}}(\beta)$, possibly in the asymptotic limit  of an infinitely long interaction time $t \rightarrow \infty$.  Under this premise,  
since no external work contributes to the
energy balance,  the heat absorbed by A 
can be legitimately identified with the  increases of the local energy  of A, i.e. ~\cite{Vinjanampathy2016,FINITE}
\begin{eqnarray} \label{DEFQA} 
\Delta Q_{\rm A} = {\mbox{Tr}} [ \hat{H}_{{\rm A}} (\hat{\rho}_{\rm A}(t) - \hat{\rho}_{\rm A}(0))].\end{eqnarray} 
%\red{
%\sout{The conservation of  the total energy  in the model, i.e. ${\mbox{Tr}}[\hat{H}(\hat{\rho}_{\rm AB}(t) - \hat{\rho}_{\rm AB}(0))]=0$, 
%implies that this quantity must be equal to minus the term ${\mbox{Tr}}[\hat{H}_{\rm B} (\hat{\rho}_{\rm B}(t) - \hat{\rho}_{\rm B}(0))]$ that  in the same spirit 
% of the interpretation of $\Delta Q_{\rm A}$ can be seen as the heat $\Delta Q_{\rm B}$ absorbed by the bath in the process,  minus a  contribution }
 %\begin{eqnarray} \sout{\Delta E_{\rm int} :={\mbox{Tr}}[\hat{H}_{\rm int} (\hat{\rho}_{\rm AB}(t) - \hat{\rho}_{\rm AB}(0))]\;,}\label{DEFDE} 
% \end{eqnarray}  \sout{that instead 
% has a less clear operational meaning. }}
The conservation of  the total energy  in the model, i.e. ${\mbox{Tr}}[\hat{H}(\hat{\rho}_{\rm AB}(t) - \hat{\rho}_{\rm AB}(0))]=0$, implies that
\begin{equation}
\Delta Q_{\rm A}= - \Delta Q_{\rm B} - \Delta E_{\rm int}
\end{equation}
where $\Delta Q_{\rm B}:={\mbox{Tr}}[\hat{H}_{\rm B} (\hat{\rho}_{\rm B}(t) - \hat{\rho}_{\rm B}(0))]$ is the heat absorbed by the bath in the process, while the energy contribution
\begin{eqnarray} 
 \Delta E_{\rm int} :={\mbox{Tr}}[\hat{H}_{\rm int} (\hat{\rho}_{\rm AB}(t) - \hat{\rho}_{\rm AB}(0))]\;,\label{DEFDE} 
 \end{eqnarray}
holds a less clear operational interpretation.
 In the weak-coupling regime, the latter term is typically  neglected either because assumed to be small as compared to $\Delta Q_{\rm A,B}$  or, more formally,  by simply considering 
interaction Hamiltonians  that only exchanges excitations between A and B (i.e. $[\hat{H}_{\rm int}, \hat{H}_{\rm A} + \hat{H}_{\rm B}]=0$), leading 
  to the identity 
$\Delta Q_{\rm A} =-  \Delta Q_{\rm B}$.
The Clausius inequality can then be derived by 
interpreting  the l.h.s. term of~(\ref{eq:Clausius})   as the  variation 
  of the von Neumann entropy 
computed on the initial state $\hat{\rho}_{\rm A}(0)$ and its evolved counterpart $\hat{\rho}_{\rm A}(t)$,
i.e. $\Delta S_A = S(\hat{\rho}_{\rm A}(t))-S(\hat{\rho}_{\rm A}(0))$ with $S(\hat{\rho}_{\rm A}(t)):= -{\mbox{Tr}}[ \hat{\rho}_{\rm A}(t) \ln \hat{\rho}_{\rm A}(t)]$. Accordingly we can now write 
\begin{eqnarray}\label{IMPOA} \Delta S_{\rm A} - \beta \Delta Q_{\rm A} = S(\hat{\rho}_{\rm A}(0)\|\hat{\eta}^{(\beta)}_{\rm A}) - S(\hat{\rho}_{\rm A}(t)\|\hat{\eta}^{(\beta)}_{\rm A}),\end{eqnarray}  
where 
 $S(\hat{\rho}_{\rm A}\|\hat{\eta}^{(\beta)}_{\rm A}):= {\mbox{Tr}}[\hat{\rho}_{\rm A} ( \ln \hat{\rho}_{\rm A} - \ln \hat{\eta}^{(\beta)}_{\rm A})]$ is the relative entropy~\cite{HOLEVOBOOK} of the density matrices $\hat{\rho}_{\rm A}$ and  $\hat{\eta}^{(\beta)}_{\rm A}$.
 Assuming the invariance of $\hat{\eta}^{(\beta)}_{\rm A}$ under $\Phi_{0\rightarrow t}$ (which given our working hypothesis should hold at least in the asymptotic limit of $t\rightarrow \infty$), the inequality (\ref{eq:Clausius})  then follows as a consequence of the monotonicity property of the relative entropy under CPT transformations~\cite{BreuerPetruccione}, i.e.  
 \begin{eqnarray} S(\hat{\rho}_{\rm A}(t)\|\hat{\eta}^{(\beta)}_{\rm A})&=& 
 S(\Phi_{0\rightarrow t}[ \hat{\rho}_{\rm A}(0)]\|\Phi_{0\rightarrow t}[  \hat{\eta}^{(\beta)}_{\rm A}])\nonumber \\
 & \leq &S(\hat{\rho}_{\rm A}(0)\|\hat{\eta}^{(\beta)}_{\rm A})\;. \label{IMPO55} \end{eqnarray} Equation~(\ref{IMPOA}) also clarifies that, at least in the asymptotic limit where
 $\hat{\rho}_{\rm A}(t)$ reaches thermalization,   the Clausius bound will in general not be tight, as in this case Eq.~(\ref{IMPOA}) reduces 
 to 
 \begin{eqnarray}\label{IMPOAa} \Delta S_{\rm A} - \beta \Delta Q_{\rm A} \Big|_{t \rightarrow  \infty}= S(\hat{\rho}_{\rm A}(0)\|\hat{\eta}^{(\beta)}_{\rm A})\;,\end{eqnarray}
 which due to the positivity of the relative entropy is not null, unless $\hat{\rho}_{\rm A}(0)$ was already at equilibrium.  
On the other hand, the extrinsic bound~(\ref{eq:Info}) admits an even simpler derivation: it follows as a consequence 
of  the sub-additivity property of the von Neumann entropy and its invariance under unitary transformations~\cite{HOLEVOBOOK}, 
i.e. $S(\hat{\rho}_{\rm A}(t))+S(\hat{\rho}_{\rm B}(t))\geq S(\hat{\rho}_{\rm AB}(t))=S(\hat{\rho}_{\rm AB}(0))=S(\hat{\rho}_{\rm A}(0))+S(\hat{\rho}_{\rm B}(0))$, or equivalently, from the positivity of the quantum mutual information~\cite{HOLEVOBOOK} 
\begin{eqnarray} 
\label{MUTUAL}
0 &\leq& I_{{\rm A:B}} (t):= S(\hat{\rho}_{\rm A}(t))+S(\hat{\rho}_{\rm B}(t))-S(\hat{\rho}_{\rm AB}(t)),\end{eqnarray} 
with $\hat{\rho}_{\rm B}(t)=\mbox{Tr}_{\rm A}[ \hat{\rho}_{\rm AB}(t)]$ the reduced density matrix of the bath at time $t$.
As anticipated~(\ref{eq:Clausius}) and (\ref{eq:Info})  are not independent: following the derivation  of 
Ref.~\cite{Reeb2014} it is possible to show that the difference between their right-hand-side terms satisfies the identity
\begin{equation}
    \label{eq:Wolf}
    \beta \Delta Q_{\rm A} + \Delta S_{\rm B} = - S(\hat{\rho}_{\rm B}(t)  || \hat{\eta}^{(\beta)}_{\rm B}) - \beta \Delta E_{\rm int}\;,
\end{equation}
(indeed by invoking the energy conservation identity  $\Delta Q_{\rm B} = -\Delta Q_{\rm A}  - \Delta E_{\rm int}$ and exploiting the connection between  
${I}_{\rm A: B}(t)$, $\Delta S_{\rm A}$ and $\Delta  S_{\rm B}$ detailed in Eq.~(\ref{MUTUAL}), the identity 
~(\ref{eq:Wolf}) reduces to the expression 
   $\beta \Delta Q_{\rm B} = - \Delta S_{\rm A} + {I}_{\rm A: B}(t)  + S(\hat{\rho}_{\rm B}(t) || \hat{\eta}_{\rm B}^{(\beta)})$
of~\cite{Reeb2014}).
Notice that for general choices of the system/bath coupling the term on the r.h.s. of Eq.~(\ref{eq:Wolf}) does not have a definite sign with the notable exceptions of those models for which 
$\Delta E_{\rm int}$ is exactly zero (e.g. the case where $\hat{H}_{\rm int}$ commutes with $\hat{H}_{\rm A}$ and $\hat{H}_{\rm B}$): under this circumstance the positivity of the relative entropy
ensures that $\beta \Delta Q_{\rm A}$  is smaller than $-\Delta S_{\rm B}$ making the extrinsic bound~(\ref{eq:Info}) tighter than (\ref{eq:Clausius}).

\subsection{Markovian regime} 
Let us focus on the case of a thermalizing processes described by  CPT-families of maps $\{ \Phi_{0\rightarrow t}\}$  which are time-homogeneous and  Markovian~\cite{BREUER}. 
 As they
 keep no record of the initial time  (i.e. ${\Phi_{0\rightarrow t}}= {\Phi_{t}}$)  and  obey a semigroup property
(i.e. ${\Phi_{t+\tau}}= {\Phi_{t}}\circ {\Phi_{\tau}}$ for all $t,\tau\geq 0$, with "$\circ$" indicating the composition of superoperator), the corresponding dynamics can be expressed in terms of 
 a master equation~\cite{BreuerPetruccione} with a Gorini-Kossakowsky-Sudarshan-Lindblad (GKSL) generator~\cite{GKSL1,GKSL2} formed by a local Hamiltonian contribution $\hat{H}_{\rm{A}}$ and a purely dissipative term  which 
 effectively accounts for the influence of the bath B. Under these assumptions the derivation of the Clausius inequality given in the previous section 
 can be generalized to bound the differential entropy increase
 $\partial  S_{\rm {A}}(t) := S(\hat{\rho}_{\rm A}(t+dt))- S(\hat{\rho}_{\rm A}(t))$  at a generic time $t\geq0$ of the temporal evolution of A in terms of the corresponding differential heat increment $\partial Q_{\rm A}(t): =  {\mbox{Tr}} [ \hat{H}_{\rm{A}} (\hat{\rho}_{\rm A}(t+dt) - \hat{\rho}_{\rm A}(t))]$, i.e. 
  \begin{eqnarray}  \label{eq:Clausiusdiff}
\partial S_{\rm{A}}(t)  \geq  \beta \partial Q_{\rm A}(t) \;.
\end{eqnarray} 
This supersedes the finite time interval version of the bound which can be now derived from  (\ref{eq:Clausiusdiff}) by direct integration, 
and it 
implies that, for time-homogeneous Markovian processes, the gap between the r.h.s. and the l.h.s. of the Clausius inequality is 
a non decreasing function of time.
   An analogous treatment of  Eq.~(\ref{eq:Info}) is more problematic as the Hamiltonian derivation of GKSL 
master equation relays on special assumptions  on the AB couplings that hide the modifications induced on the bath degrees of freedom. 
For properly accounting these effects we need  a framework that permits  to treat all the degrees of freedom, including the bath ones, on an equal footing, e.g. exploiting the CM approach we analyze in the following section. 

\section{Thermalization  via collisions} \label{SEC2}  
%%%%%%%%%%%%%%%%%%%%%%%%%%%%%%%
 \begin{figure*}[!t]
 \centering
 \includegraphics[width=\textwidth]{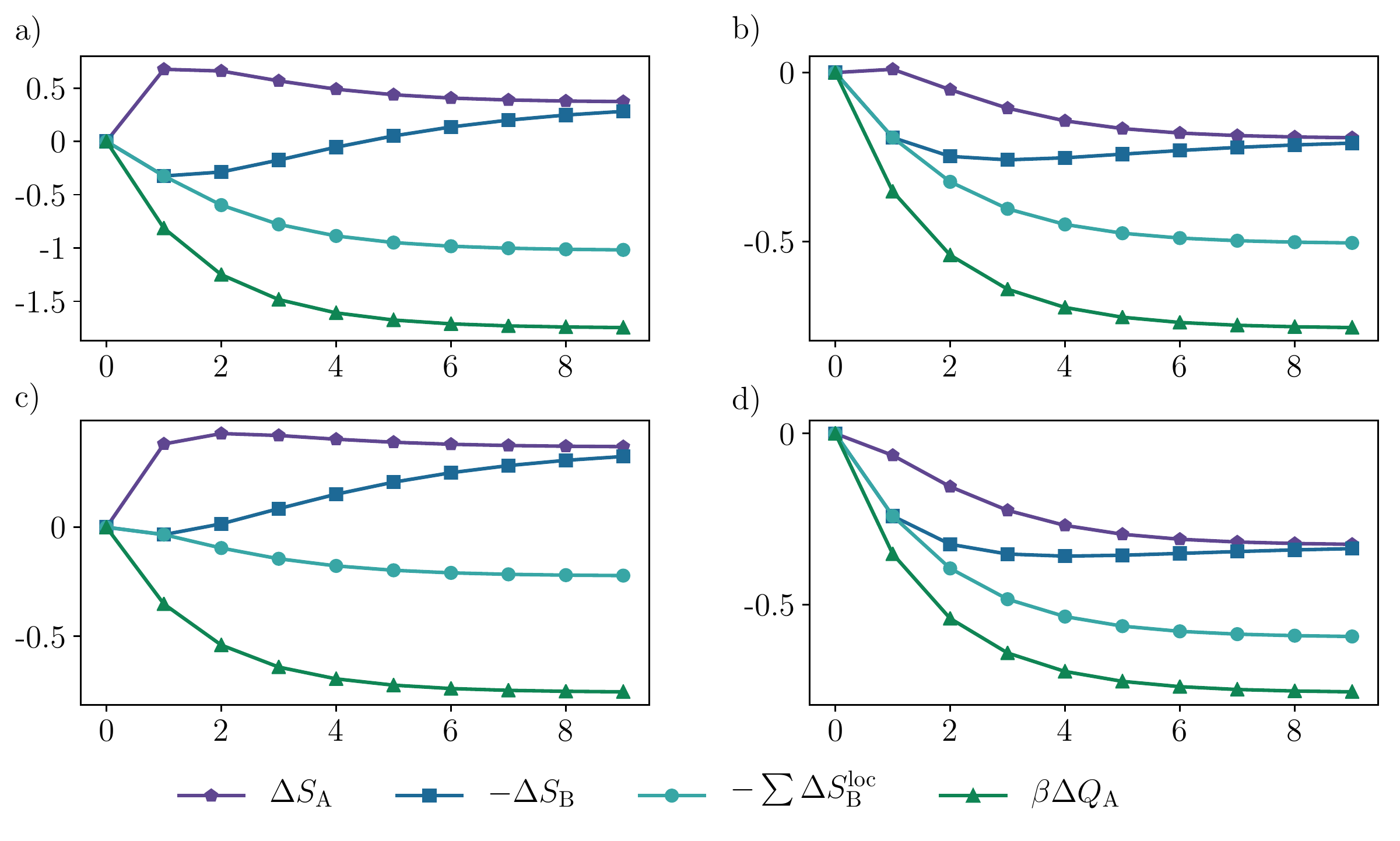}
  \caption{(color online) Numerical evaluation of the bounds for the entropy production in the CM for a qubit system. In the plot the behavior of
 $\Delta S^{(n)}_{\rm A}$ and of the quantities $\beta\Delta Q^{(n)}_{\rm A}$ (intrinsic bound, Eq.~(\ref{eq:Clausius111})), $-\Delta S^{(n)}_{\rm B}$ (extrinsic bound, Eq.~(\ref{eq:Info111})), and 
 $-\Delta S_{\rm B}^{(n,loc)}$ (local version of the extrinsic bound, Eq.~(\ref{eq:InfoNEasdW10}))  are shown as a function of the collisional step index $n$ for four different initial states of A represented
 by the Bloch vectors $\vec{r}^{(0)} =(0,0,1)$ (ground state of A, panel a), $(1/2,0,0)$ (panel b), $(1,0,0)$ (panel c), and  $(0,0,0)$ (completely mixed state of A, panel d), respectively. 
 The reported values fulfill the ordering anticipated in  Eq~(\ref{eq:ordering}),  
$-\Delta S^{(n)}_{\rm B}$ always providing  the optimal bound.    
 Furthermore  as $n$ increases,  $-\Delta S^{(n)}_{\rm B}$ approaches $\Delta S^{(n)}_{\rm A}$ saturating the extrinsic bound~\eqref{eq:Info111} in agreement with the asymptotic factorization conjecture of Eq.~(\ref{RISSCAR111}).
In all the plots we set $\beta=1$ and $\theta=0.75$ (i.e. below the threshold $\arctan2$, see Sec.~\ref{FACTSEC}).}
 \label{fig:plot_sep}
 \end{figure*}
 
%%%%%%%%%%%%%%%%%%%%%%%%%%%%%%%%%%%%%%%%%%

 CMs are simplified yet effective descriptions of the AB interactions~\cite{first_coll_mod,coll_therm,COLL2,COLL3}.  Here 
 the thermal bath B  is depicted as a many-body quantum system formed by a huge collection of $N$ (possibly infinite)  identical  and non-interacting, small subsystems ${\rm b}_1$, ${\rm b}_2$, $\cdots$, ${\rm b}_N$, characterized by local Hamiltonians $\hat{H}_{{\rm b}_1}$, $\hat{H}_{{\rm b}_2}$, $\cdots$, $\hat{H}_{{\rm b}_N}$
and  initialized  in the same Gibbs thermal state $\hat{\eta}^{(\beta)}_{\rm b}:= e^{-\beta \hat{H}_{\rm b}}/Z_{\rm{b}}(\beta)$. Such subsystems interact sequentially with A for a finite time $\delta t$ and are then discarded
 to enforce Markovianity~\cite{BreuerPetruccione,Man2018,Campbell2018,Scarani2002}. 
 Within this model the 
 global state of the joint system AB after the first $n$ collisions is hence expressed as 
 \begin{equation}
    \label{eq:evolved state R}
    \hat{\rho}_{\rm AB}^{(n)} =  { \cal{U}}_{n} \circ \cdots \circ  {\cal {U}}_{2}\circ  {\cal  {U}}_{1} \left[ \hat{\rho}_{A}^{(0)} \otimes (\hat{\eta}_{\rm b}^{(\beta)})^{\otimes N} \right] \;,
\end{equation} 
with  $\hat{\rho}_{A}^{(0)}$ and $(\hat{\eta}_{\rm b}^{(\beta)})^{\otimes N}$ being respectively the initial density matrices of A and B, and where 
for $k$ an integer, ${\cal U}_{k}(\cdots) = \hat{U}_k (\cdots) \hat{U}_k^\dag$ is a unitary conjugation induced by the interaction between A and the $k$-th bath subsystem  ${\rm b}_k$.
Accordingly A will now evolve via a stroboscopic  sequence of jumps in which, at the $n$-th step,  
 the reduced density matrix $\hat{\rho}_{\rm A}^{(n-1)}:=\mbox{Tr}_{\rm B} [  \hat{\rho}_{\rm AB}^{(n-1)} ]$ gets mapped  into 
 \begin{eqnarray} 
 \hat{\rho}_{\rm A}^{(n)}\label{DEF111} 
=   \mbox{Tr}_{{\rm b}} [ {\cal{U}}_n (\hat{\rho}_{\rm A}^{(n-1)}\otimes \hat{\eta}_{{\rm b}}^{(\beta)} )]=: \Phi[ \hat{\rho}_{\rm A}^{(n-1)}]\;, 
\end{eqnarray} 
where we decomposed the partial trace upon B into a sequence of partial traces upon the various bath subsystems to remove all the degrees of freedom
but the $n$-th ones. 
Equation~(\ref{DEF111})  makes explicit 
the Markovian character of the evolution and, by iteration, clarifies that in the CM the elements of the family $\{ \Phi_{t}\}_{t\geq0}$
are replaced by the powers of the map $\Phi$, i.e. $\hat{\rho}_{\rm A}^{(n)} = \Phi^{n}[ \hat{\rho}_{\rm A}^{(0)}]$. 

In order to ensure that for large enough $n$ the system A will reach the thermal equilibrium state $\hat{\eta}_{\rm A}^{(\beta)}$, i.e.
 \begin{eqnarray} \lim_{n\rightarrow \infty} \hat{\rho}_{\rm A}^{(n)} = \hat{\eta}^{(\beta)}_{\rm A}\;, \label{RISSCAR} 
 \end{eqnarray} 
 we follow Ref.~\cite{Scarani2002,SWAP1,SWAP2,SWAP3,Lorenzo2015a,Campbell2018}  assuming  the  environment subsystems ${\rm b}_{\rm{k}}$ to be
isomorphic with A, identifying their local Hamiltonians with the one of A (i.e.  $\hat{H}_{\rm A}\equiv \hat{H}_{{\rm b}_k}$), and taking the unitaries that mediate the collisions   to be  partial swap operators. 
Specifically for all $n$ integer we assume 
\begin{equation}
    \label{eq:partial swap}
    \hat{U}_{n} = \exp[i  \theta  \hat{\mathbb{S}}_n] = \cos \theta  \, \mathbb{I}_n + i \sin\theta \, \hat{\mathbb{S}}_n\;,
\end{equation}
where $\theta\in]-\pi, \pi]$ is a dimensionless parameter that gauges the  strength of the collisional event, and which is proportional 
to the collisional time $\delta t$, while 
 $\hat{\mathbb{S}}_n=\hat{\mathbb{S}}_n^\dag=\hat{\mathbb{S}}_n^{-1}$ is the swap operator coupling system A with the $n$-th environmental bath subsystem (when acting on states of the form $|\psi\rangle_{\rm A} \otimes |\phi\rangle_{{\rm b}_n}$ it exchanges them, i.e.  $\hat{\mathbb{S}}_n |\psi\rangle_{\rm A} \otimes |\phi\rangle_{{\rm b}_n} = |\phi\rangle_{\rm A} \otimes |\psi\rangle_{{\rm b}_n}$). 
Besides implying the property (\ref{RISSCAR}) for all input $\hat{\rho}_{\rm A}^{(0)}$ (this being true~\cite{Scarani2002}  as long as  
$\theta$ is not an integer  multiple of $\pi$),
 the choices detailed above ensure  that the sum $\hat{H}_{\rm A} + \sum_{n=1}^N \hat{H}_{{\rm b}_n}$ of the local Hamiltonians of A and B commutes with the 
 unitary transformation $\hat{U}_{n} \cdots  \hat{U}_{2} \hat{U}_{1}$. Therefore, similarly  to the $\Delta E_{\rm int}=0$ case of the Hamiltonian model,  the energy variations of A in the CM  are compensated by 
an opposite variation for the bath, leading to the identity 
 $\Delta Q^{(n)}_{\rm A} = - \Delta Q^{(n)}_{\rm B}$, where for ${\rm X}= {\rm A},{\rm B}$,
 $\Delta Q_{\rm X}^{(n)}:= {\mbox{Tr}} [ \hat{H}_{\rm X} (\hat{\rho}_{\rm AB}^{(n)} - \hat{\rho}_{\rm AB}^{(0)})]$ 
measures the heat absorbed by the ${\rm X}$ system 
between the zero-th and $n$-th step of the process.
Using the same  arguments adopted in the 
 previous section on the state (\ref{eq:evolved state R}) 
 it is also possible to show that   both the Clausius inequality~(\ref{eq:Clausius}) and the extrinsic bound (\ref{eq:Info}) still hold true, i.e.  \begin{eqnarray}  \label{eq:Clausius111}
\Delta S_{\rm A}^{(n)} &\geq&  \beta \Delta Q_{\rm A}^{(n)}\;,  \\ 
 \Delta S_{\rm A}^{(n)} &\geq& - \Delta S_{\rm B}^{(n)}\;,  \label{eq:Info111}
\end{eqnarray} 
with  $\Delta S_{\rm A}^{(n)} :=S(\hat{\rho}_{\rm A}^{(n)})-S(\hat{\rho}_{\rm A}^{(0)})$ 
and $\Delta S_{\rm B}^{(n)}:= S(\hat{\rho}_{\rm B}^{(n)})-S(\hat{\rho}_{\rm B}^{(0)})$
 gauging the entropy  differences between the zero-th and $n$-th step of the process. A detailed derivation of this
 relations is provided in the Appendix. 
 %%%%%%%%%%%%%%% 
    \begin{figure*}[!t]
        \centering
        % \includegraphics[scale=0.63]{mut_info_mix_state}
        %\includegraphics[scale=0.5]{mut_info_pure_state}
        %\subfloat[a][Plot of $I(A:B)$ and $\Delta S_A-\beta\Delta Q$ for $\hat{\rho}(0)=\frac{1}{2}(\dyad{0}+\dyad{1})+\frac{1}{4}(\dyad{0}{1}+\dyad{1}{0})$ and $\hat{\eta}^{(\beta)}=0.12\dyad{0}+0.88\dyad{1}$.]
     %   {\includegraphics[scale=0.5]{partial_swap_rand_state_w_inset}\label{fig:mut_info_mix}}%\\
        %\subfloat[b][Same as the other one, but with logarithmic scaling.]
        %{\includegraphics[scale=0.3]{partial_swap_rand_state_log}\label{fig:mut_info_pure}}
%        \subfloat[c][Same as the other one, but with logarithmic scaling.]
\quad
{\includegraphics[width=\textwidth]{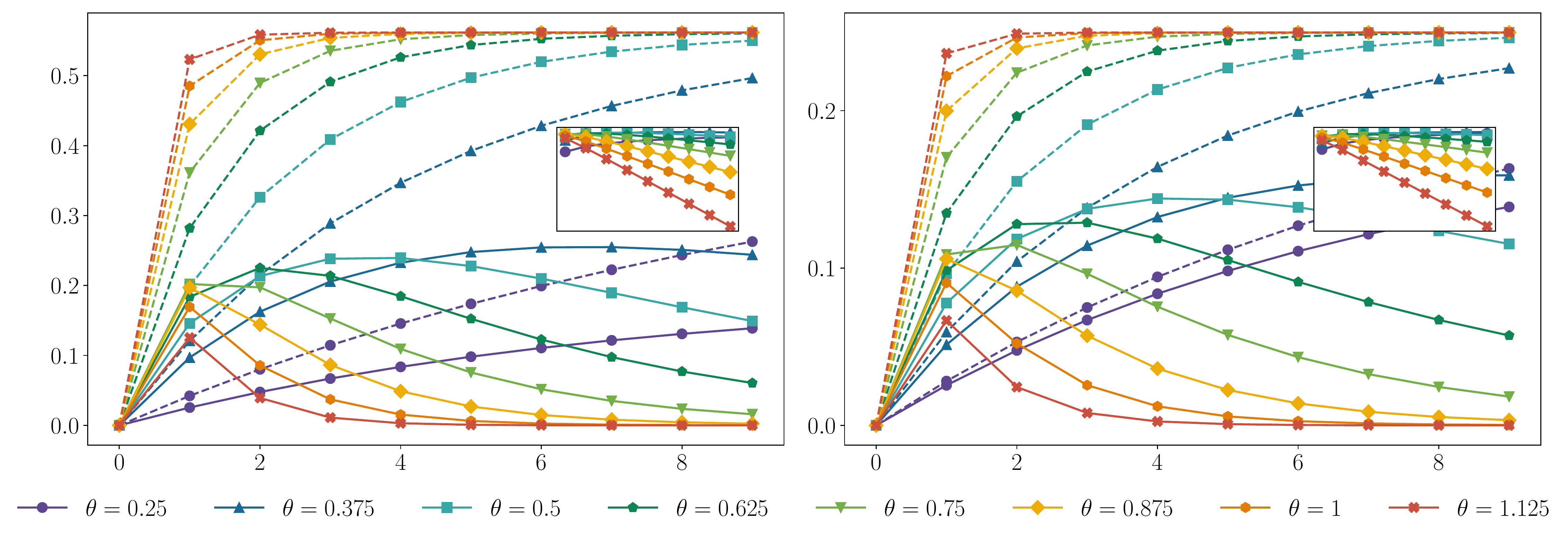}\label{fig:mut_info_pure}}
        \caption{ 
        %{\bf the plots here are roughly the correct ones, the parameters will change, the colors will change and the plot on the right will most probably be an inset. }\\
        (color online) Numerical evaluation of the intrinsic~(\ref{eq:Clausius111}) and extrinsic~(\ref{eq:Info111}) bounds for the entropy production in the CM for a qubit system.
Dashed lines show the behavior of $\Delta S_{\rm A}^{(n)} - \beta \Delta Q_{\rm A}^{(n)}$  as a function of the collisional step $n$  for  various values of the partial swap parameter $\theta$ of Eq.~(\ref{eq:partial swap}). Notice that as expected for Markovian processes this is an increasing functions of  $n$ which saturates  to the value 
 $S(\hat{\rho}_{\rm A}^{(0)}\|\hat{\eta}^{(\beta)}_{\rm A})$ in agreement with~(\ref{IMPOAa}). The continuous lines represent instead the
behavior of $\Delta S_{\rm A}^{(n)} + \Delta S_{\rm B}^{(n)}$. In this case the function has no monotonic behavior and as $n$ increases it tends to nullify in agreement with  the factorization prediction of Eqs.~(\ref{eq:Info111222}) and (\ref{RISSCAR111}). In the inset the same quantity is shown in logarithmic scale.
   In all plots  the input state of A is associated to $\vec{r}^{(0)}=(1/2,0,0),$ 
   while the temperature of the bath is such that $\beta = 1$ in the left panel and $\beta=0.5$ in the right panel. 
        }        \label{fig:sec_princ}
\end{figure*}
%%%%%%%%%%%%%%%%%%%%%%%%%%%%%%%%%
 Here we only stress that  in computing the term on the r.h.s. of (\ref{eq:Info111})
 we treat B as a real many-body quantum system, retaining all inter-particle correlations which may arise 
 between its sub-environment components  ${\rm b}_1$, ${\rm b}_2$, $\cdots$,  ${\rm b}_N$ due to their collisions
 with A. According to this choice the explicit evaluation of $\Delta S_{\rm B}^{(n)}$ is a highly demanding task as it
 requires the full diagonalization of the density matrix $\hat{\rho}_{\rm B}^{(n)}$ of the environment. 
 An alternative, and much easier to compute,  formulation of this inequality could also be obtained by effectively removing such correlations (e.g. by replacing  the joint
 state of AB that is emerging from the $n$-th collision with its reduced factorized counterpart~\cite{Campbell2018}):
this yields the following "local" version of the extrinsic bound 
  \begin{eqnarray}  
 \Delta S_{\rm A}^{(n)} &\geq&   - \Delta S_{\rm B}^{(n,loc)} \label{eq:InfoNEasdW10} \;, 
 \end{eqnarray} 
  which is provably weaker than the one we present in Eq.~(\ref{eq:Info111}) -- 
 see Appendix~\ref{APPENDIXINC} for details.
 Another important observation is the fact that 
for the CM we analyze here, the identity  Eq.~(\ref{eq:Wolf})  still applies.  Due to the
 absence of the  $\Delta E_{\rm int}$ contribution (see discussion above),  this implies that 
 for the present model the extrinsic bound~(\ref{eq:Info111}) and its local counterpart (\ref{eq:InfoNEasdW10}),
 always beat the intrinsic one leading to the following ordering 
   \begin{eqnarray}  
 \Delta S_{\rm A}^{(n)} \geq   - \Delta S_{\rm B}^{(n)}  \geq  - \Delta S_{\rm B}^{(n,loc)}  \geq  \beta \Delta Q_{\rm A}^{(n)} \label{eq:ordering} \;. 
 \end{eqnarray} 
 
As a matter of fact it turns out that  the inequality (\ref{eq:Info111}) is asymptotically optimal, the gap between   $\Delta S_{\rm A}^{(n)}$ and $-\Delta S_{\rm B}^{(n)}$ being
exponentially decreasing. Introducing the quantum mutual information $ I^{(n)}_{{\rm A}:{\rm B}}:= S(\hat{\rho}_{\rm A}^{(n)})+ S(\hat{\rho}_{\rm B}^{(n)})-S(\hat{\rho}_{\rm AB}^{(n)})
$~\cite{HOLEVOBOOK} and noticing that $S(\hat{\rho}_{\rm AB}^{(n)})= S(\hat{\rho}_{\rm AB}^{(0)})=S(\hat{\rho}_{\rm A}^{(0)})+S(\hat{\rho}_{\rm B}^{(0)})$ due to unitary invariance of the von Neumann entropy, 
this result can be equivalently  casted in terms of the following identity 
   \begin{eqnarray} 
 \lim_{n\rightarrow \infty} 
 I^{(n)}_{{\rm A}:{\rm B}}=0 \label{eq:Info111222}\;. \end{eqnarray}
 The above expression is a major improvement with respect to the identity~(\ref{RISSCAR}): while the latter  implies that the global state of A and B approaches the Gibbs configuration $\hat{\eta}^{(\beta)}_{\rm A}$ locally on A, Eq.~(\ref{eq:Info111222}) imposes explicit  factorization  of the  joint state 
  in the asymptotic limit of large $n$, i.e. 
 \begin{eqnarray} \lim_{n\rightarrow \infty} \hat{\rho}_{\rm AB}^{(n)} = \hat{\eta}^{(\beta)}_{\rm A}\otimes {\hat \Lambda}_{\rm B}  \;, 
  \label{RISSCAR111} 
 \end{eqnarray}  
  the density matrix   ${\hat \Lambda}_{\rm B}$ maintaining  a functional dependence upon the input state $\hat{\rho}_{\rm A}^{(0)}$ of A as a consequence of the unitary mapping~(\ref{eq:evolved state R}).  
For  $|\theta|\geq \arctan 2\simeq   1.107$ (strong-collisions regime) 
an explicit proof the asymptotic factorization property (\ref{eq:Info111222}) is presented in Sec.~\ref{SUBSECNEW}: it works for any finite dimensional system A and for all choices of the inverse temperature $\beta$.
For lower values of $|\theta|$ our argument fail, still we conjecture that~(\ref{RISSCAR111})
should hold also in those cases. In support to this conjecture we present 
some numerical evidences where we have tested the model for the case where
A and the subsystems ${\rm b}_n$ are qubits with local Hamiltonians   $\hat{H}_{\rm A}=\frac{1}{2}\hat{\sigma}_{\rm A}^{(3)}$ and $\hat{H}_{\rm b} =\frac{1}{2}\hat{\sigma}_{\rm b}^{(3)}$, $\hat{\sigma}^{(3)}$  being the third Pauli operator and the energy scale being measured in units $\hbar \omega =1$. Results are reported in Sec.~\ref{SUBSECNEW2} and summarized 
in Figs.~\ref{fig:plot_sep} and~\ref{fig:sec_princ}.
% where, for both the intrinsic and extrinsic bounds
%~\eqref{eq:Clausius111} and~\eqref{eq:Info111}.
% we plot the differences between the associated  l.h.s. and r.h.s. terms as a function of $n$ for various values of the coupling parameter $\theta$ and for
 %some choices of the input state of A. 

\subsection{The qubit case} \label{SUBSECNEW2} 
Here we study in details the CM  for the special case where A and the sub-environments ${\rm b}$ are qubit systems. 

Adopting the Bloch sphere representation we write the state $\hat{\rho}_{\rm A}^{(n)}$ of A after the $n$-th collision 
\begin{eqnarray}
\label{eq:vec_qubit}
\hat{\rho}_{\rm A}^{(n)}=\frac{\hat{\mathbb{I}}_{\rm A}+\vec{r}^{(n)}\cdot\vec{{\sigma}}_{\rm A}}{2}\;, 
\end{eqnarray}
with $\hat{\mathbb{I}}_{\rm A}$ and $\vec{\sigma}_{\rm A} = (\hat{\sigma}^{(1)}_{\rm A},\hat{\sigma}^{(2)}_{\rm A},\hat{\sigma}^{(3)}_{\rm A})$ being respectively the identity  and Pauli vector operators.
From this we can compute the associated entropy $S(\hat{\rho}_{\rm A}^{(n)})$ and the mean energy of the state as
\begin{eqnarray} 
S(\hat{\rho}_{\rm A}^{(n)}) &=& H_2\left(\frac{1+ |\vec{r}^{(n)}|}2\right)\;, \\
E_{\rm A}^{(n)} &:=& \mbox{Tr}[ \hat{H}_{\rm A} \hat{\rho}_{\rm A}^{(n)}] = r_3^{(n)}/2\;, 
\end{eqnarray} 
where  $H_2(x):= -x \ln x - (1-x) \ln (1-x)$ is the Shannon binary entropy functional and where we have used the fact that $\hat{H}_{\rm A}=\frac{1}{2} \hat{\sigma}^{(3)}_{\rm A}$. Accordingly, introducing the Bloch vector $\vec{r}^{(0)}$ of the input state 
$\hat{\rho}_{\rm A}^{(0)}$ of A we can then write 
\begin{eqnarray} \label{EXP0} 
\Delta S_{\rm A}^{(n)} &=& H_2\left(\frac{1+ |\vec{r}^{(n)}|}2\right)-H_2\left(\frac{1+ |\vec{r}^{(0)}|}2\right) \;,  \\ 
\Delta Q_{\rm A}^{(n)} &=&  (r_3^{(n)}-r_3^{(0)})/2\;.\label{EXP00} 
\end{eqnarray} 
From  the definition of $H_2(x)$ it follows that $\Delta S_{\rm A}^{(n)}$ is an increasing function of 
the length $|\vec{r}^{(n)}|$ of the Bloch vector $\vec{r}^{(n)}$, while $\Delta Q_{\rm A}^{(n)}$ is just linearly dependent upon the
$z$-axis component of such vector. 

A closed expression for these quantities can then be obtained by exploiting  the properties of the partial swap transformation (\ref{eq:partial swap}), to recast (\ref{DEF111}) into the following recursive mapping 
\begin{eqnarray}
\vec{r}^{(n)}=\cos^{2}\theta \;  \vec{r}^{(n-1)}+\sin^{2}\theta \;\vec{s}\;,  \label{QUESTA}
\end{eqnarray}
with $\vec{s}$ being the Bloch vector associated with the input state  $\hat{\eta}_{\rm b}^{(\beta)}=e^{-\beta \hat{\sigma}^{(3)}_{\rm b}/2}/Z_{\rm{b}}(\beta)$ of the 
environmental subsystem ${\rm b}$, i.e. 
\begin{eqnarray} 
\vec{s} = (0,0,s)\;, \qquad  s:=s(\beta) = - \tanh(\beta/2) \;. \end{eqnarray} 
Iterating, expression~(\ref{QUESTA}) can be formally integrated giving 
\begin{eqnarray}
\vec{r}^{(n)}&=&\cos^{2n}\theta\; \vec{r}^{(0)}+(1-\cos^{2n}\theta)\; \vec{s} \nonumber \\
&=&  \vec{s} + \cos^{2n}\theta \Delta \vec{r}^{(0)} \label{VECTR} \;,\end{eqnarray}
with $\Delta \vec{r}^{(0)} :=  \vec{r}^{(0)} -\vec{s}$ the difference between the Bloch vector $\vec{r}^{(0)}$ of the input state 
of  A 
and $\vec{s}$. 
The length and the $z$-axis component  of the vector~(\ref{VECTR})  can then be computed as 
\begin{eqnarray}
|\vec{r}^{(n)}|
&=&\sqrt{s^2 + \cos^{4n}\theta |\Delta \vec{r}^{(0)}|^2 + \cos^{2n}\theta \; s \; \Delta {r}_3^{(0)} }\;,\nonumber \\  \\ 
{r}_{3}^{(n)} -{r}_3^{(0)}
&=& (\cos^{2n}\theta -1)\; \Delta {r}_3^{(0)} \;. 
\end{eqnarray} 

By the same token we can now compute the local output states of the environmental subsystems $\hat{\rho}^{(n)}_{{\rm b}}=\mbox{Tr}_{\rm A} [ \hat{U} (\hat{\rho}_{\rm A}^{(n-1)}\otimes \hat{\eta}_{{\rm b}}^{(\beta)} )\hat{U}^\dag ]
$ which
via Eq.~(\ref{eq:InfoNEasdW}) provide a lower bound to $\Delta S_{\rm A}^{(n)}$. Following the same derivation given above
the Bloch vector of such state can be expressed as
\begin{eqnarray} \nonumber 
 \vec{s}^{(n)}&=&\sin^{2}\theta \;  \vec{r}^{(n-1)}+\cos^{2}\theta \;\vec{s} \\ &=&
\vec{s} + \sin^{2}\theta  \cos^{2(n-1)}\theta \Delta \vec{r}^{(0)}\;,
  \label{QUESTAQUI}
\end{eqnarray}
where the last identity follows from (\ref{VECTR}). The length of such vector is hence 
\begin{eqnarray}
|\vec{s}^{(n)}| \nonumber 
&=&\left(s^2 + \sin^{4}\theta \cos^{4(n-1)}\theta |\Delta \vec{r}^{(0)}|^2  \right. \\
&&\left. + \sin^{2}\theta \cos^{2(n-1)}\theta \; s \; \Delta {r}_3^{(0)} \right)^{1/2} \;,
\end{eqnarray} 
yielding 
\begin{eqnarray} \label{EXP1} 
\Delta S_{\rm B}^{(loc)}= % \sum_{k=1}^n S(\hat{\rho}^{(k)}_{{\rm b}}) -n S(\hat{\eta}_{\rm b}^{(\beta)})- = \\
  \sum_{k=1}^n  H_2\left(\frac{1+ |\vec{s}^{(k)}|}2\right) -n H_2\left(\frac{1+ s}2\right),
\end{eqnarray}
 for the quantity (\ref{eq:InfoNEasdW}) of the Appendix that define  the local version~(\ref{eq:InfoNEasdW10}) 
of the extrinsic bound. 

Expressions~(\ref{EXP0}), (\ref{EXP00}), and (\ref{EXP1}) are used for the plots of Figs.~\ref{fig:plot_sep}
and ~\ref{fig:sec_princ}. The evaluation of $\Delta S_{\rm B}^{(n)}$ instead requires a complete diagonalization of the many-body quantum state of the environment B.

%%%%%%%%%%%%%%%%%%%%%%%%%%%%%%%%%%%%%%%%%%%%%%%%%%%

\section{Correlations decay}\label{SUBSECNEW}
This section  focuses on the factorization  property~(\ref{RISSCAR111}).
As a preliminary observation we notice that, irrespectively of the values of $\theta$ and $\beta$, Eq.~(\ref{RISSCAR111})
is trivially verified when  the input state of A is already at thermal equilibrium, i.e. $\hat{\rho}_{\rm A}^{(0)} =\hat{\eta}^{(\beta)}_{\rm A}$. Indeed  under this condition one has that for all $n$ one has 
\begin{eqnarray} 
 \hat{\rho}_{\rm AB}^{(n)} = \hat{\eta}^{(\beta)}_{\rm A}\otimes  (\hat{\eta}_{\rm b}^{(\beta)})^{\otimes N} \;, 
\end{eqnarray} 
due to the fact that given $\hat{\mathbb{S}}$ the swap operator acting on the Hilbert space ${\cal H}^{\otimes 2}$, the state  $\hat{\eta}\otimes \hat{\eta}$ commutes with it (i.e.  $[\hat{\mathbb{S}}, \hat{\eta}\otimes \hat{\eta}] = 0$),  and hence with $e^{i \theta \hat{\mathbb{S}}}$, thus leading to 
 $e^{i \theta \hat{\mathbb{S}}} (\hat{\eta}\otimes \hat{\eta}) e^{- i \theta \hat{\mathbb{S}}} = \hat{\eta}\otimes \hat{\eta}$.
 Also, Eq.~(\ref{RISSCAR111}) can be easily shown to hold for  arbitrary inputs, in the case where the bath temperature
 is zero ($\beta \rightarrow \infty$) and the local Hamiltonians of the model  have a non degenerate ground state $|0\rangle$.
 In this case in fact the Gibbs states $\hat{\eta}_{\rm b}^{(\beta)}$ correspond to the pure vectors $|0\rangle_{\rm b}$ 
 while (\ref{RISSCAR}) yields
 \begin{eqnarray} \lim_{n\rightarrow \infty} \hat{\rho}_{\rm A}^{(n)} = |0\rangle_{\rm A}\langle 0| \;, \label{RISSCARzero} 
 \end{eqnarray} 
which can only be fulfilled by having a joint state $\hat{\rho}_{\rm AB}^{(n)}$ that asymptotically approaches
a state of the form $|0\rangle_{\rm A}\langle 0| \otimes {\hat \Lambda}_{\rm B}$. 

In Sec.~\ref{FACTSEC} we shall prove that~(\ref{RISSCAR111}) holds for the CM model as long as the strength of the partial swapping is sufficiently large. 
In Sec.~\ref{APPECM}, instead the factorization the property~(\ref{RISSCAR111}) is shown to hold for a slightly modification
of the scheme where we do alternate sequences of collisions with a full dephasing process on A.

%%%%%%%%%%%%%%%%%%%%%%%%%%%%%%%%%%%%%

\subsection{Factorization proof for the strong-collisions regime}  \label{FACTSEC} 

Consider now the non trivial case where $\hat{\rho}_{\rm A}^{(0)} \neq \hat{\eta}^{(\beta)}_{\rm A}$ and $\beta$ is finite. 
In order to prove the identity~(\ref{RISSCAR111}) we notice that  Eq.~(\ref{eq:evolved state R}) which defines the joint state of AB 
after $n$ collisions implies the following 
recursive formula 
     $\hat{\rho}_{\rm AB}^{(n)} =    {\cal {U}}_n [ \hat{\rho}_{\rm AB}^{(n-1)}]$. 
We now express such state as the sum of two terms 
\begin{equation} 
\hat{\rho}_{\rm AB}^{(n)} = \hat{R}_{\rm AB}^{(n)} +  \hat{T}_{\rm AB}^{(n)}\;,
\end{equation}
with $\hat{R}_{\rm AB}^{(n)}$ representing the contribution where A factorizes from B and is in
 $\hat{\eta}^{(\beta)}_{\rm A}$ (i.e. $\hat{R}_{\rm AB}^{(n)} = \hat{\eta}^{(\beta)}_{\rm A} \otimes \hat{R}_{\rm B}^{(n)}$),
 while $\hat{T}_{\rm AB}^{(n)}$ containing all the remaining ones. An explicit derivation 
 of such decomposition can be formally derived via the following construction: 
  for $n=0$, using the fact that $\hat{\rho}_{\rm A}^{(0)} \neq \hat{\eta}^{(\beta)}_{\rm A}$
   we set 
    $\hat{R}_{\rm AB}^{(0)} = 0$ and 
    $\hat{T}_{\rm AB}^{(0)}=\hat{\rho}_{\rm AB}^{(0)}$. Then for  $n=1$ we use 
    Eq.~(\ref{eq:partial swap}) and the properties of the swap operator to write  
\begin{eqnarray} 
  \hat{R}_{\rm AB}^{(1)} &=& \sin^2\theta \left( \hat{\mathbb{S}}_1 \hat{\rho}_{\rm AB}^{(0)} \hat{\mathbb{S}}_1 \right) \nonumber  \\
   &=& \sin^2\theta \; \left(\hat{\eta}_{\rm A}^{(\beta)}  \otimes \hat{\rho}_{{\rm b}_1}^{(0)}  \otimes [\hat{\eta}_{\rm b}^{(\beta)}]^{\otimes N-1}\right) \;,\\
\hat{T}_{\rm AB}^{(1)} &=& \cos^2\theta \; \hat{\rho}_{\rm AB}^{(0)} + i \sin\theta \cos\theta \left[ \hat{\mathbb{S}}_1 , \hat{\rho}_{\rm AB}^{(0)}\right] \;,
\end{eqnarray} 
where $[\cdots, \cdots]$ stands for the commutator. 
For higher values of $n$ we can derive a recursive formula connecting $\hat{R}_{\rm AB}^{(n+1)}$, $\hat{T}_{\rm AB}^{(n+1)}$ to $\hat{R}_{\rm AB}^{(n)}$, $ \hat{T}_{\rm AB}^{(n)}$, by noticing that 
\begin{equation} 
    {\cal {U}}_{n+1}[ \hat{R}_{\rm AB}^{(n)} ] =\hat{R}_{\rm AB}^{(n)} \;,
\end{equation}
which follows once more from the fact that 
 states $\hat{\eta}\otimes \hat{\eta}$ are invariant under partial swaps (see argument at the beginning of the section). Therefore, the only part of ${\cal{U}}_{n+1}  [\hat{T}_{\rm AB}^{(n)}]$ that contributes to $\hat{R}_{\rm AB}^{(n+1)}$ is the one which has $\hat{\mathbb{S}}_{n+1}$ either only on the right and or only on  the left. Accordingly we have
\begin{align} 
    \hat{R}_{\rm AB}^{(n+1)} &=  \hat{R}_{\rm AB}^{(n)} + \sin^2 \theta\;   \hat{\mathbb{S}}_{n+1} \;  \hat{T}_{\rm AB}^{(n)}\; \hat{\mathbb{S}}_{n+1}\;,\\
    \hat{T}_{\rm AB}^{(n+1)} &= \cos^2\theta  \hat{T}_{\rm AB}^{(n)}  + i \sin\theta \cos\theta\;  \left[ \hat{\mathbb{S}}_{n+1} ,  \hat{T}_{\rm AB}^{(n)}\right]\;.
\end{align} 
Exploiting the subadditivity of the norm we then get 
\begin{eqnarray} 
    \|\hat{T}_{\rm AB}^{(n+1)}\| &\leq&  | \cos^2\theta | \|  \hat{T}_{\rm AB}^{(n)} \| + |\sin\theta \cos\theta| \left\|  \left[ \hat{\mathbb{S}}_{n+1} ,  \hat{T}_{\rm AB}^{(n)}\right]  \right\| \nonumber \\ 
    &\leq& ( | \cos^2\theta | + 2  |\sin\theta \cos\theta|  )\;  \|  \hat{T}_{\rm AB}^{(n)} \| \;,
\end{eqnarray} 
where in the second line we used the fact that $ \hat{\mathbb{S}}_{n+1}$ is unitary to claim that  $\|  \hat{\mathbb{S}}_{n+1}  \hat{T}_{\rm AB}^{(n)} \| = \|  \hat{T}_{\rm AB}^{(n)}\|$ (the above results being true in any operator norm).  
 Iterating this we can then write 
\begin{align}
    \label{DD} 
    \|\hat{T}_{\rm AB}^{(n+1)}\| &\leq (  | \cos^2\theta | + 2  |\sin\theta \cos\theta|  )^{n+1} \;  \| \hat{\rho}_{\rm AB}^{(0)} \| \;,
\end{align}
where we used  the fact that  $\hat{T}_{\rm AB}^{(0)}=\hat{\rho}_{\rm AB}^{(0)}$. Now we observe that for 
\begin{eqnarray} |\theta| >\arctan 2\label{critical}\;, \end{eqnarray} we have 
  $| \cos^2\theta | + 2  |\sin\theta \cos\theta|   < 1$
  and hence from (\ref{DD}) 
  \begin{equation} 
  \lim_{n\rightarrow \infty} \| \hat{T}_{\rm AB}^{(n)}\| =0\;,
  \end{equation} 
  implying that in the large $n$ limit $\hat{T}_{\rm AB}^{(n)}$ approaches zero allowing us to identify 
  $\hat{\rho}_{\rm AB}^{(n)}$ with $\hat{R}_{\rm AB}^{(n)}$ as required by Eq.~(\ref{RISSCAR111}).

\subsection{Asymptotic factorization for CM with a little help from  full dephasing on A} \label{APPECM} 

Here we show that the asymptotic factorization  property %by showing that 
~(\ref{RISSCAR111}) can be proven under a slight modification of the CM where,
instead of letting A and B evolving under a sequence of collisional events as in (\ref{eq:evolved state R})
every $k\gg1$ collisions we force A to undergo 
 full dephasing transformation ${\cal D}_{\rm A}$ which destroys  all its off-diagonal elements with respect to the
local energy eigenbasis $\{ |j\rangle_{\rm A}\}$, i.e. 
\begin{eqnarray} \label{FULL}  
{\cal D}_{\rm A}[ |j\rangle_{\rm A} \langle j'|] = \delta_{j,j'} |j\rangle_{\rm A} \langle j|\;,
\end{eqnarray} 
with $\delta_{j,j'}$ the Kronecker delta.
For the sake of simplicity we present this argument for the special case of A being a qubit, but the same can be generalized to
arbitrary dimensions. Also we stress that, as long as the property (\ref{RISSCAR}) is verified together with the assumption that the state of $\hat{\eta}_{\rm A}^{(\beta)}$ will
not evolve during a collisional event, the derivation we present below does not relay on the specific form of the  unitaries 
given in Eq.~(\ref{eq:partial swap}).

Let us hence divide the subsystems of B into groups of $k$ elements: the set  ${\rm B}_1$, containing the first $k$ subenvironments, the set ${\rm B}_2$ containing the second $k$ subenvironments, and so on and so forth. 
Take then the joint state of A and ${\rm B}_1$ after the first $k$ unitary collisions have been performed, and  expand it with respect to the local energy basis of 
A, i.e. 
\begin{eqnarray} 
 \hat{\rho}_{\rm AB_1}^{(k)} &=&  { \cal{U}}_{k} \circ \cdots \circ  {\cal {U}}_{2}\circ  {\cal  {U}}_{1} \left[ \hat{\rho}_{\rm A}^{(0)} \otimes (\hat{\eta}_{\rm b}^{(\beta)})^{\otimes k} \right]  \nonumber \\ 
&=& \sum_{j, j'} |j\rangle_{\rm A} \langle j'|\otimes \hat{\Pi}_{\rm B_1}^{(j,j')}(\rho_{\rm A}^{(0)}) \;, \label{EVk} 
 \end{eqnarray} 
with $\hat{\Pi}_{\rm B_1}^{(j,j')}(\rho_{\rm A}^{(0)}) := {_{\rm A}\langle} j'|  \hat{\rho}_{\rm AB}^{(k)}  |j\rangle_{\rm A}$ being
operators of ${\rm B}_1$  which inherit a linear dependence upon the input state $\hat{\rho}_{\rm A}^{(0)}$ of A. 
Taking the partial trace with respect to ${\rm B}_1$  this yields the density operator
\begin{eqnarray} 
\hat{\rho}_{\rm A}^{(k)} &=& \sum_{j, j'} M^{(k)}_{j,j'}(\rho_{\rm A}^{(0)})  |j\rangle_{\rm A} \langle j'|
\end{eqnarray} 
with matrix coefficients 
\begin{eqnarray} 
M^{(k)}_{j,j'}(\rho_{\rm A}^{(0)})  := \mbox{Tr}_{{\rm B}_1 } [\hat{\Pi}_{{\rm B}_1}^{(j,j')}(\rho_{\rm A}^{(0)}) ] \;.
\end{eqnarray}
From (\ref{RISSCAR}) we know that for large $k$ should approach the stationary configuration $\hat{\eta}_{\rm A}^{(\beta)}$, which by construction, is diagonal with respect to the basis $\{ |j\rangle_{\rm A}\}$ with eigenvalues equal to
$\eta_j^{(\beta)}  = e^{- \beta \hbar E_j}/Z(\beta)$.  Accordingly, for any given positive $\epsilon <1$, we can choose $k$ sufficiently large to guarantee that
the following inequalities hold, 
\begin{eqnarray} 
|M^{(k)}_{j,j}(\rho_{\rm A}^{(0)}) - \eta_j^{(\beta)}|&<& \epsilon\;, \label{ACCD} 
\\
 |M^{(k)}_{j,j'}(\rho_{\rm A}^{(0)}) |&<& \epsilon\;, \quad \forall j\neq j'\;. 
\end{eqnarray} 
If we next apply the full dephasing (\ref{FULL})  to the  state (\ref{EVk}) we get: 
\begin{eqnarray} 
&& \hat{\rho}_{\rm AB}^{(k)} \rightarrow
 |0\rangle_{\rm A} \langle 0|\otimes \hat{\Pi}_{\rm B_1}^{(0,0)}(\rho_{\rm A}^{(0)}) + |1\rangle_{\rm A} \langle 1|\otimes \hat{\Pi}_{\rm B_1}^{(1,1)}(\rho_{\rm A}^{(0)}) \nonumber \;, \\ 
  \label{EVk2} 
 \end{eqnarray} 
where, for the first time, we explicitly used the fact that system A is a qubit.
Summing and subtracting the term $\tfrac{\eta_0^{(\beta)}}{\eta_1^{(\beta)}} 
 \hat{\Pi}_{\rm B_1}^{(1,1)}(\rho_{\rm A}^{(0)})$ and using the definition of $\hat{\eta}_{\rm A}^{(\beta)}$ the r.h.s. of this  expression can thus  be rewritten as
 \begin{eqnarray} 
|0\rangle_{\rm A} \langle 0|\otimes  \hat{\Delta}^{(k)}_{\rm B_1} (\rho_{\rm A}^{(0)})
+
   \hat{\eta}_{\rm A}^{(\beta)}
 \otimes\hat{\Xi}_{\rm B_1}(\rho_{\rm A}^{(0)}) \;, 
  \label{EVk2} 
 \end{eqnarray} 
 with $\hat{\Xi}_{\rm B_1}(\rho_{\rm A}^{(0)})  := \hat{\Pi}_{\rm B_1}^{(1,1)}(\rho_{\rm A}^{(0)})  / \eta_1^{(\beta)}$ and with 
 \begin{eqnarray}  \label{DDSF}
 \hat{\Delta}^{(k)}_{\rm B_1} (\rho_{\rm A}^{(0)}):= \hat{\Pi}_{\rm B_1}^{(0,0)}(\rho_{\rm A}^{(0)}) - \tfrac{\eta_0^{(\beta)}}{\eta_1^{(\beta)}} 
 \hat{\Pi}_{\rm B_1}^{(1,1)}(\rho_{\rm A}^{(0)}) \;,
 \end{eqnarray} 
an operator whose trace norm  $\|\hat{\Delta}^{(k)}_{\rm B_1} (\rho_{\rm A}^{(0)})\|_1$, for sufficiently  large $k$,  can be forced to be strictly smaller than one
thanks to (\ref{ACCD}). Indeed summing and subtracting  $\tfrac{M^{(k)}_{0,0}(\rho_{\rm A}^{(0)})}{M^{(k)}_{1,1}(\rho_{\rm A}^{(0)})}       
\hat{\Pi}_{\rm B_1}^{(1,1)}(\rho_{\rm A}^{(0)})$ to $\hat{\Delta}^{(k)}_{\rm B_1} (\rho_{\rm A}^{(0)})$ and using the 
 triangular inequality of the trace norm we can write 
$\|\hat{\Delta}^{(k)}_{\rm B_1} (\rho_{\rm A}^{(0)})\|_1 \leq \alpha^{(k)} + \beta^{(k)}$ 
with  
\begin{eqnarray} 
\alpha^{(k)} &:=&  M^{(k)}_{0,0}(\rho_{\rm A}^{(0)})
 \left\|     \tfrac{\hat{\Pi}_{\rm B_1}^{(0,0)}(\rho_{\rm A}^{(0)})}{M^{(k)}_{0,0}(\rho_{\rm A}^{(0)}} -  \tfrac{\hat{\Pi}_{\rm B_1}^{(1,1)}(\rho_{\rm A}^{(0)})}{M^{(k)}_{1,1}(\rho_{\rm A}^{(0)}}\right\|_1\nonumber \\
\beta^{(k)} &:=& 
 \left| {M^{(k)}_{0,0}(\rho_{\rm A}^{(0)})}   -  \tfrac{\eta_0^{(\beta)}}{\eta_1^{(\beta)}} {M^{(k)}_{1,1}(\rho_{\rm A}^{(0)})}  \right| \left\|  \tfrac{\hat{\Pi}_{\rm B_1}^{(1,1)}(\rho_{\rm A}^{(0)})}{M^{(k)}_{1,1}(\rho_{\rm A}^{(0)}}\right\|_1 \;.\nonumber 
\end{eqnarray}
The thesis then follows by noticing that for sufficiently large $k$,  we can ensure that  $\alpha^{(k)}$ is a quantity smaller than $1$ with  $\beta^{(k)}$  being arbitrary small. Regarding $\alpha^{(k)}$ 
this can be shown by exploiting the fact  that   in the limit of high $k$,
 $M^{(k)}_{0,0}(\rho_{\rm A}^{(0)})$ approaches $\eta_0^{(\beta)}$ which for $\beta>0$ is always strictly smaller than one, while, 
since $\tfrac{\hat{\Pi}_{\rm B_1}^{(0,0)}(\rho_{\rm A}^{(0)})}{M^{(k)}_{0,0}(\rho_{\rm A}^{(0)}}$ and $\tfrac{\hat{\Pi}_{\rm B_1}^{(1,1)}(\rho_{\rm A}^{(0)})}{M^{(k)}_{1,1}(\rho_{\rm A}^{(0)}}$ are
properly 
normalized density matrices the norm of their difference is certainly smaller than or equal to  one. 
 Regarding $\beta^{(k)}$ instead we can use 
 (\ref{ACCD}) to show that $ \left| {M^{(k)}_{0,0}(\rho_{\rm A}^{(0)})}   -  \tfrac{\eta_0^{(\beta)}}{\eta_1^{(\beta)}} {M^{(k)}_{1,1}(\rho_{\rm A}^{(0)})}  \right|$ approach zero 
 for large values of $k$, while $\left\|  \tfrac{\hat{\Pi}_{\rm B_1}^{(1,1)}(\rho_{\rm A}^{(0)})}{M^{(k)}_{1,1}(\rho_{\rm A}^{(0)}}\right\|_1=1$ because it is the trace norm of a properly normalized state. 

Equation~(\ref{EVk2}) represents the state of the ${\rm A{B}_1}$ after $k$ unitary collisions and a single dephasing event ${\cal D}_{\rm A}$. 
We now  repeat  the full procedure introducing the second $k$ subsystems of B, i.e.  the elements of the subset ${\rm B}_2$. We notice the part of the state ~(\ref{EVk2}) which has A already in  $\hat{\eta}_{\rm A}^{(\beta)}$
does not evolve. The only element that undergoes to modification is the first component of the state. Iterating the above procedure we hence 
arrive to 
\begin{eqnarray} 
|0\rangle_{\rm A} \langle 0|\otimes  \hat{\Delta}^{(k)}_{\rm B_1} (\rho_{\rm A}^{(0)})\otimes \hat{\Delta}^{(k)}_{\rm B_2} (|0\rangle_{\rm A})
+
   \hat{\eta}_{\rm A}^{(\beta)}
 \otimes \hat{\Xi}_{\rm B_1B_2}(\rho_{\rm A}^{(0)})  \;, \nonumber 
  \label{EVk2332} 
 \end{eqnarray} 
 with $\hat{\Delta}^{(k)}_{\rm B_2} (|0\rangle_{\rm A})$ as in (\ref{DDSF}) for $\rho_{\rm A}^{(0)}= |0\rangle_{\rm A} \langle 0|$ and $\hat{\Xi}_{\rm B_1B_2}(\rho_{\rm A}^{(0)})$
 a proper operator of ${\rm B_1B_2}$. 
 By the same token after $q$ of such steps we get 
 \begin{eqnarray} 
|0\rangle_{\rm A} \langle 0|\otimes  \hat{\Delta}^{(k)}_{\rm B_1} (\rho_{\rm A}^{(0)})\otimes_{\ell=2}^q  \left(\hat{\Delta}^{(k)}_{\rm B_\ell} (|0\rangle_{\rm A})\right)
\nonumber \\ +
   \hat{\eta}_{\rm A}^{(\beta)}
 \otimes \hat{\Xi}_{\rm B_1B_2\cdots B_q}(\rho_{\rm A}^{(0)})  \;. \nonumber 
  \label{EVk2332sdf} 
 \end{eqnarray} 
 Notice that the first contribution has a trace norm which is  equal to 
$\| |0\rangle_{\rm A} \langle 0|_1\otimes  \hat{\Delta}^{(k)}_{\rm B_1} (\rho_{\rm A}^{(0)})\otimes_{\ell=2}^q  \left(\hat{\Delta}^{(k)}_{\rm B_\ell} (|0\rangle_{\rm A})\right)\|_1\nonumber \\
=\| \hat{\Delta}^{(k)}_{\rm B_1} (\rho_{\rm A}^{(0)})\| \; \| \hat{\Delta}^{(k)}_{\rm B_\ell} (|0\rangle_{\rm A})\|_1^{q-1}$ and hence is exponentially decreasing in $q$. Accordingly we can claim that for large $q$ the state of AB  will  be determined by the second contribution which 
explicitly factorizes as in Eq.~(\ref{RISSCAR111}).

%%%%%%%%%%%%%%%%%%%%%%%%%%%%%%%%%%%%%%%%%
\section{Conclusions}\label{SEC3} 
The scheme of Ref.~\cite{Scarani2002} is arguably the simplest thermalization model one can analyze which, within the assumptions of the CM approach, appears to be consistent 
both  thermodynamically and from the point of view of open quantum dynamics. 
Our analysis clarifies that in this context  the Clausius inequality is always outperformed by the extrinsic bound that relates
 $\Delta S_{\rm A}$ to the entropy increment of the thermal environment B.
 
 Most interesting, in the limit of infinitely many collisions, the latter turns out to be asymptotically
 optimal, indicating that the model induces a complete factorization of A from B.
To our understanding, this progressive factorization arises as the result of the balance of two competing effects that take place at each swapping collision: on one hand every interaction with a new ancilla tends, in principle, to establish new correlations between the environment and the system. On the other hand, this same interaction tends to reduce the correlations the system established with the previous ancillas, by transforming them into intra-environement correlations via partial replacement of the system degree of freedom with those of the new ancilla due to the action of the swap gate. As a matter of fact the first mechanism becomes more and more feeble approaching the fixed point of the evolution. 

We have shown that such asymptotic factorization
 holds true at least when the strength of the collision is sufficient large: also in view of our numerical analysis, we suspect however that this result should be fairly general and we
 plan to further investigate it in a next future. 

%%%%%%%%%%%%%%%%%%%%%%%%%%%%%%%%%%%%%%%%%%

\acknowledgements
A.D.P acknowledges the financial support from the University of Florence in the framework of the University Strategic Project Program 2015 (project BRS00215).

\appendix

%\section{Proof of Eq. (\ref{eq:Wolf})}
%
%Invoking the energy conservation identity  $\Delta Q_{\rm B} = -\Delta Q_{\rm A}  - \Delta E_{\rm int}$ and exploiting the connection between  
%${I}_{\rm A: B}(t)$ and $\Delta S_{\rm A}$ and $\Delta  S_{\rm B}$ detailed in Eq.~(\ref{MUTUAL}), the identity 
%~(\ref{eq:Wolf}) reduces to the expression 
%\begin{equation} 
%    \label{eq1}
%   \beta \Delta Q_{\rm B} = - \Delta S_{\rm A} + {I}_{\rm A: B}(t)  + S(\hat{\rho}_{\rm B}(t) || \hat{\eta}_{\rm B}^{(\beta)}),
%\end{equation}
%which in Ref.~\cite{Reeb2014} was shown to apply for the Hamiltonian AB interaction model.
%

\section{Explicit Derivation of the entropic bounds for the CM} \label{APPENDIXINC} 
The  global form~(\ref{eq:Clausius111}) of the Clausius inequality is obtained by applying the relative-entropy monotonicity argument
to the states $\hat{\rho}_{\rm A}^{(n)}$ and $\hat{\eta}_{\rm A}^{(\beta)}$. The incremental version of this follows instead by
using the same procedure by comparing the entropies of $\hat{\rho}_{\rm A}^{(n)}$ and $\hat{\rho}_{\rm A}^{(n-1)}$ via Eq.~(\ref{DEF111}), obtaining the inequality 
 \begin{eqnarray} d S_{\rm A}^{(n)} \geq  \beta d Q_{\rm A}^{(n)} \label{inutile}\;,  \end{eqnarray} with 
     $d S_{\rm A}^{(n)}:=S(\hat{\rho}_{\rm A}^{(n)})-S(\hat{\rho}_{\rm A}^{(n-1)})$,  $d Q_{\rm A}^{(n)}:= {\mbox{Tr}} [ \hat{H}_A (\hat{\rho}_{\rm A}^{(n)} - \hat{\rho}_{\rm A}^{(n-1)})]$.

The extrinsic bound~(\ref{eq:Info111}) is obtained by invoking the sub-additivity of the von Neumann entropy of the density matrix~(\ref{eq:evolved state R}): here    $\Delta S_{\rm B}^{(n)}$ represents 
the global entropy gain of the multipartite bath B, which properly accounts for all possible correlations between its 
 constituents ${\rm b}_1$, ${\rm b}_2$, $\cdots, {\rm b}_N$. Writing it explicitly it results in the following expression:
  \begin{eqnarray} 
 \Delta S_{\rm A}^{(n)} &\geq& - \Delta S_{\rm B}^{(n)}= n S(\hat{\eta}_{\rm b}^{(\beta)})-  S(\hat{\rho}^{(n)}_{{B}})
 \;,  \label{eq:Info111EXP}
\end{eqnarray} 
where we used the fact that at the beginning of the A--B interactions the bath is described by the factorized state where all its constituents are initialized into the same Gibbs state  $\hat{\eta}_{\rm b}^{(\beta)}$. 
An incremental version of this inequality  instead 
follows by using the same technique applied to the state~(\ref{DEF111}), i.e. 
   \begin{eqnarray}  
 d S_{\rm A}^{(n)} \geq - d S_{\rm b}^{(n)} =
    S(\hat{\eta}_{\rm b}^{(\beta)})-S(\hat{\rho}^{(n)}_{{\rm b}}) \label{eq:Info1112} \;.
 \end{eqnarray}  
Here $d S_{\rm b}^{(n)}$  represents the local entropy variation of the $n$-th environmental subsystem ${\rm b}_n$ after its collision with A (by construction such system
  evolves from $\hat{\eta}_{\rm b}^{(\beta)}$
  to
$\hat{\rho}^{(n)}_{{\rm b}}:=\mbox{Tr}_{\rm A} [ \hat{U} (\hat{\rho}_{\rm A}^{(n-1)}\otimes \hat{\eta}_{{\rm b}}^{(\beta)} )\hat{U}^\dag ]$).
   It is worth stressing that, at variance with the intrinsic bound where Eq.~(\ref{eq:Clausius111}) can be seen as a consequence of Eq.~(\ref{inutile}) via direct summation of the latter over all 
  collisions, (\ref{eq:Info1112}) results in a weaker bound for the global entropy production of A than Eq.~(\ref{eq:Info111}). Indeed 
  by summing over the first $n$ collisions Eq.~(\ref{eq:Info1112}) yields the inequality~(\ref{eq:InfoNEasdW10}) with 
    \begin{eqnarray}  
 %\Delta S_{\rm A}^{(n)} &\geq&   - \Delta S_{\rm B}^{(n,loc)} \label{eq:InfoNEasdW110} \;, \\ 
  \Delta S_{\rm B}^{(n,loc)}&:=&\sum_{k=1}^n S(\hat{\rho}^{(k)}_{{\rm b}}) -n S(\hat{\eta}_{\rm b}^{(\beta)}), \label{eq:InfoNEasdW} 
 \end{eqnarray}  
 the bound being outperformed by~(\ref{eq:Info111EXP}) due to entropy  subaddivitity, i.e.   $\sum_{k=1}^n S(\hat{\rho}^{(k)}_{{\rm b}})\geq S(\hat{\rho}^{(n)}_{{B}})$
 the $\hat{\rho}^{(k)}_{{\rm b}}$ being the reduced density matrix of the $k$-th ancillary system of 
 $\hat{\rho}_{\rm B}^{(n)}$. As a matter of fact, Eq.~(\ref{eq:InfoNEasdW}) is exactly the bound one would get from 
 (\ref{eq:Info111EXP}) when removing all the intra-particle correlations between the subenvironments elements,
 i.e. by replacing $\hat{\rho}^{(n)}_{{\rm b}}$ with the product state formed by the reduced density matrices of its constituents  $\hat{\rho}^{(n)}_{{\rm b}} \rightarrow \hat{\rho}^{(n)}_{{\rm b}_2} \otimes \hat{\rho}^{(n)}_{{\rm b}_1}\otimes \cdots \hat{\rho}^{(n)}_{{\rm b}_n}$. As explicitly noted in Ref.~\cite{Campbell2018} this procedure
  will not affect the dynamics of A (and hence its entropy increase), yet at the level of the extrinsic lower bound 
  gives worst performances than the one presented in Eq.~(\ref{eq:Info111EXP}).

 Finally we notice that a formal rewriting of the identity (\ref{eq:Wolf}) for the CM reads as
\begin{eqnarray}
    \label{eq:WolfCM}
    \beta \Delta Q^{(n)}_{\rm A} + \Delta S^{(n)}_{\rm B} &=& - S(\hat{\rho}^{(n)}_{\rm B} || \hat{\rho}^{(0)}_{\rm B})  \;,
\end{eqnarray}
which implies $- \Delta S^{(n)}_{\rm B} \geq \beta \Delta Q^{(n)}_{\rm A}$ and which can be directly proven by direct evaluation of the various terms by enforcing the local energy conservation identity $\Delta Q^{(n)}_{\rm A} = -\Delta Q^{(n)}_{\rm B}$
discussed in the main text. A similar identity holds for the incremental entropy variations, i.e.  
\begin{equation}
    \label{eq:WolfCMd}
    \beta d Q^{(n)}_{\rm A} + d S^{(n)}_{\rm b} = - S(\hat{\rho}^{(n)}_{\rm b} || \hat{\eta}^{(\beta)}_{\rm b}) \;,
\end{equation}
which upon summation over the collision index $n$ yields the inequality 
\begin{equation}
    \label{eq:WolfCManew}
- \Delta S^{(n,loc)}_{\rm B}   \geq  \beta \Delta Q^{(n)}_{\rm A}  \;,
\end{equation}
anticipated in Eq.~(\ref{eq:ordering}) of the main text.

\clearpage 
\setcounter{equation}{0}%
\setcounter{figure}{0}%
\setcounter{table}{0}%
\renewcommand{\thetable}{S\arabic{table}}
\renewcommand{\theequation}{S\arabic{equation}}
\renewcommand{\thefigure}{S\arabic{figure}}

%\onecolumngrid
%
%\begin{center}
%{\Large Supplemental Material}
%
%\vspace*{0.5cm}
%
%
%
%\vspace{1cm}
%
%\end{center}

\end{document}